\documentclass[aps,prl,reprint,superscriptaddress,floatfix,nofootinbib,showkeys,showpacs,preprintnumbers]{revtex4-1}
\usepackage{graphicx}
\graphicspath{{./figs/}{./}}
\usepackage{comment}

\usepackage{lineno}

\usepackage[usenames,dvipsnames,svgnames]{xcolor}

\usepackage{amsmath}
\usepackage{amssymb}
\usepackage{xspace}
\usepackage{xifthen}
\usepackage{eso-pic}

\newcommand{\ie}{i.e.\xspace}
\newcommand{\eg}{e.g.\xspace}

\newcommand{\reportnum}[2]{
  \AddToShipoutPictureBG*{%
    \AtPageUpperLeft{%
      \hspace{0.75\paperwidth}%
      \raisebox{#1\baselineskip}{%
        \makebox[0pt][l]{\textnormal{#2}}
  }}}%
}

\mathchardef\mhyphen="2D

\newcommand{\roughly}{\ensuremath{ {\sim}\,} }

\newlength{\dhatheight}

\newcommand{\unit}[1]{\ensuremath{\mathrm{\,#1}}\xspace}
\newcommand{\eV}{\unit{eV}}
\newcommand{\MeV}{\unit{MeV}}

\newcommand{\um}{\unit{\mu m}}

\newcommand{\us}{\unit{\mu s}}
\newcommand{\ms}{\unit{ms}}

\newcommand{\e}{\unit{e^{-}}}

\newcommand{\rms}{\unit{rms}}
\newcommand{\pix}{\unit{pix}}
\newcommand{\spl}{\unit{sample}}
\newcommand{\ermspix}{\e \rms/\pix}
\newcommand{\epixday}{\unit{\e \pix^{-1} \unit{day}^{-1}}}

\begin{document}

\title{
Single-electron and single-photon sensitivity with a silicon Skipper CCD}

\author{Javier Tiffenberg}
\email{javiert@fnal.gov}
\affiliation{Fermi National Accelerator Laboratory, PO Box 500, Batavia IL, 60510}

\author{Miguel Sofo-Haro}
\affiliation{Centro At{\'o}mico Bariloche, CNEA/CONICET/IB, Bariloche, Argentina}
\affiliation{Fermi National Accelerator Laboratory, PO Box 500, Batavia IL, 60510}

\author{Alex Drlica-Wagner}
\affiliation{Fermi National Accelerator Laboratory, PO Box 500, Batavia IL, 60510}

\author{Rouven Essig}
\affiliation{C.N. Yang Institute for Theoretical Physics, Stony Brook University, Stony Brook, NY 11794}

\author{Yann Guardincerri}
\thanks{Deceased January 2017.}
\affiliation{Fermi National Accelerator Laboratory, PO Box 500, Batavia IL, 60510}

\author{Steve Holland}
\affiliation{Lawrence Berkeley National Laboratory, One Cyclotron Rd, Berkeley, CA 94720}

\author{Tomer Volansky}
\affiliation{Raymond and Beverly Sackler School of Physics and Astronomy, Tel-Aviv University, Tel-Aviv 69978, Israel}

\author{Tien-Tien Yu}
\affiliation{Theoretical Physics Department, CERN, CH-1211 Geneva 23, Switzerland}

\reportnum{-1.5}{FERMILAB-PUB-17-183-AE}
\reportnum{-2.5}{YITP-SB-16-25} 
\reportnum{-3.5}{CERN-TH-2017-114}

\date{\today}
\begin{abstract}

We have developed a non-destructive readout system that uses a floating-gate amplifier on a thick, fully depleted charge coupled device (CCD) to achieve ultra-low readout noise of $0.068 \ermspix$.
This is the first time that discrete sub-electron readout noise has been achieved reproducibly over millions of pixels on a stable, large-area detector.
This allows the precise counting of the number of electrons in each pixel, ranging from pixels with 0 electrons to more than 1500 electrons. 
The resulting CCD detector is thus an ultra-sensitive calorimeter. 
It is also capable of counting single photons in the optical and near-infrared regime.
Implementing this innovative non-destructive readout system has a negligible impact on CCD design and fabrication, and there are nearly immediate scientific applications.  
As a particle detector, this CCD will have unprecedented sensitivity to low-mass dark matter particles and coherent neutrino-nucleus scattering, while future astronomical applications may include direct imaging and spectroscopy of exoplanets.
\end{abstract}

\maketitle

\section*{Introduction}

Charge-coupled devices (CCDs) have proven essential for photon and other particle detection in many industrial and scientific applications \cite[\eg][]{Boyle:1970,Amelio:1970,damerell:1981,Janesick:2001}.
For photon detection, typical CCD sensors rely on the photoelectric effect to absorb incident photons in a silicon substrate and generate electron-hole pairs~\cite{Amelio:1970}.  
Energetic photons ($E \gtrsim 10 \eV$) produce multiple electron-hole pairs allowing for energy measurement, while lower energy photons may only generate one or a few electron-hole pairs. 
Moreover, massive particles can create electron-hole pairs either by directly interacting with valence-band electrons or by scattering off of silicon nuclei.
In both regimes, precision measurements are limited by the readout noise of the CCD electronics.
Electronic readout noise is added to the video signal by the CCD output amplifier, and results in variations in the charge assigned to each pixel~\cite{Janesick:2001}.
Correlated double sampling techniques \cite{CDS:1050448} dramatically reduce high-frequency readout noise, and are essential for modern CCD operation.
However, low-frequency readout noise has remained a fundamental limitation on precision single-photon and single-electron counting in current CCDs. 

\begin{figure}[!t]
\centering
\includegraphics[width=\columnwidth]{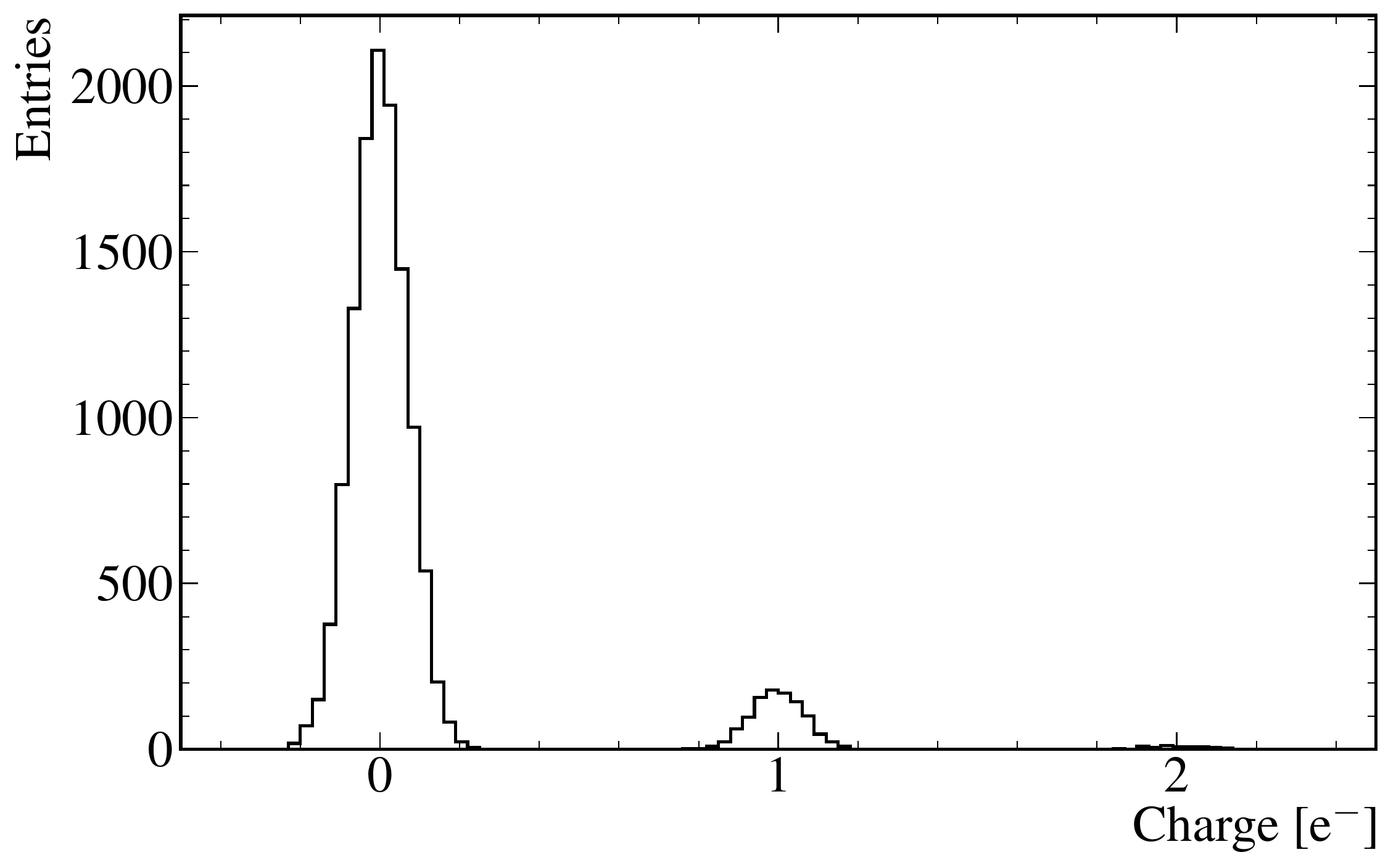}
\includegraphics[width=\columnwidth]{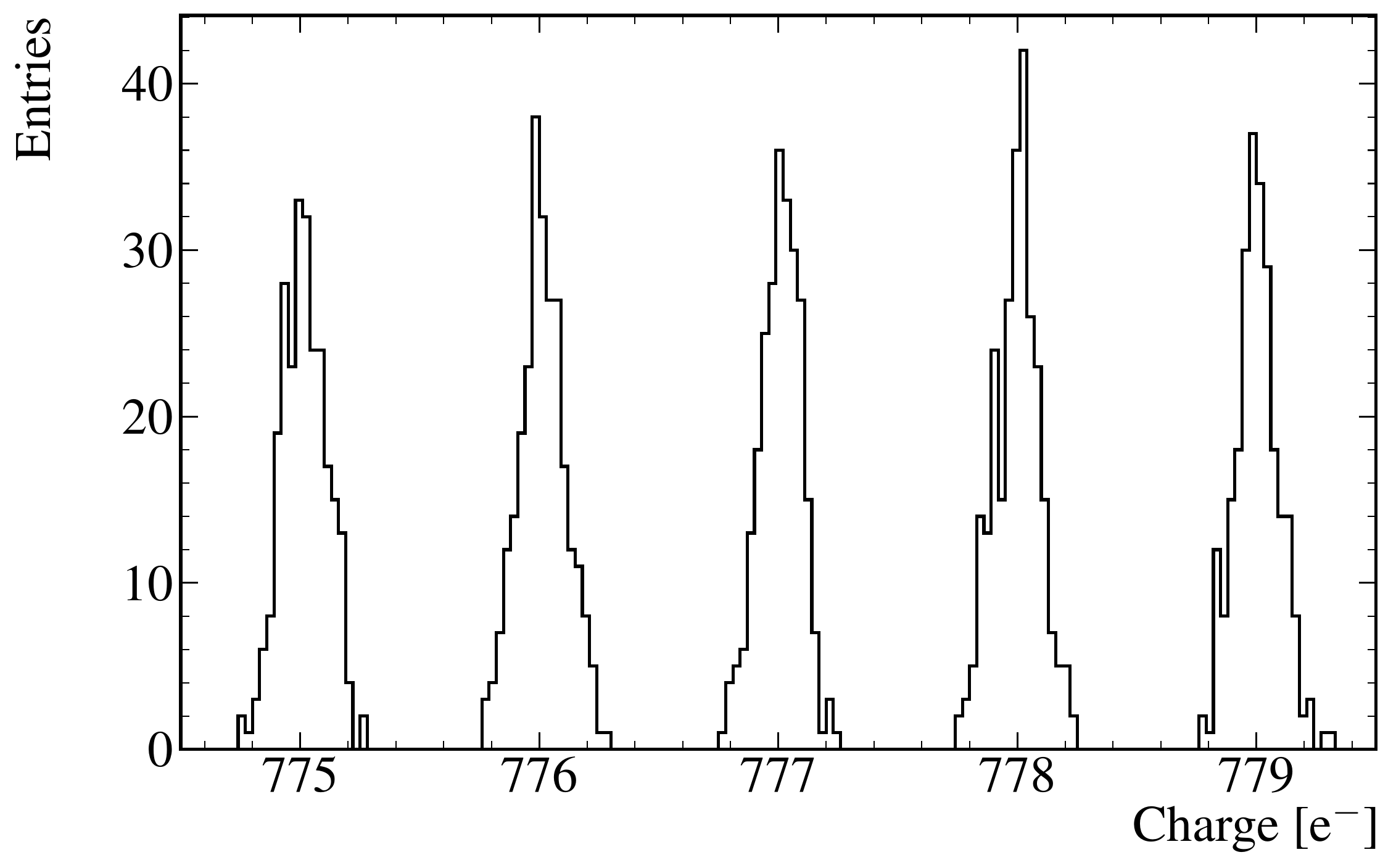}
\caption{
Single-electron charge resolution using a Skipper CCD with 4000 samples per pixel (bin width of 0.03~\e). 
The measured charge per pixel is shown for pixels with low-light level illumination ({\bf top}) and stronger illumination ({\bf bottom}).
Integer electron peaks can be distinctly resolved in both regimes contemporaneously.
The peak at 0\e has rms noise of 0.068\ermspix while the peak at 777\e has rms noise of 0.086\ermspix. 
The Gaussian fits have $\chi^2 = 22.6/22$ and $\chi^2 = 19.5/21$, respectively.  The two measurements demonstrate the single-electron sensitivity over a large dynamical range.
}
\label{fig:spectra}
\end{figure}

In conventional scientific CCDs, low-frequency readout noise results in root-mean-squared (rms) variations in the measured charge per pixel at the level of $\sim 2 \e \rms/\pix$ \cite[\eg][and references therein]{Janesick:2016,Bebek:2017}. 
\citet{Janesick:1990} proposed that low-frequency readout noise could be reduced by using a floating gate output stage \citep{Wen:1974} to perform repeated measurements of the charge in each pixel.
This multiple readout technique was implemented in the form of a ``Skipper'' CCD \cite{Janesick:1990,Chandler:1990}; however, these early detectors suffered from spurious charge generation \citep{Moroni:2012}.
Here we couple the floating gate output stage of the Skipper CCD to a small-capacitance sense node and isolate both from parasitic noise sources in order to perform multiple, independent, non-destructive measurements of the charge in a single pixel. 
The result is a drastic reduction in low-frequency readout noise to the level of $0.068 \ermspix$ (as shown in Figure~\ref{fig:spectra}).
At this noise level, the probability that the charge per pixel is mis-estimated by $>0.5\e$ is $p \sim 10^{-13}$.
This represents the first accurate single-electron counting on a large-format ($4126 \times 866$ pix) silicon detector.\footnote{In contrast, a readout noise of $0.18 \ermspix$ has been demonstrated for a $4 \times 4$ pix mini-array of silicon depleted field effect transistors \citep{Lutz:2016}.}

The low readout noise achieved by Skipper CCDs, coupled with a stable linear gain, allows charge measurement at the accuracy of individual electrons simultaneously in pixels with single electrons and thousands of electrons.
This makes the Skipper CCD the most sensitive and robust electromagnetic calorimeter that can operate at temperatures above that of liquid nitrogen.
It also allows the Skipper CCD to count individual optical and near-infrared photons.
Because non-destructive readout is achieved without any major modifications to the CCD fabrication process, this new technology can be directly implemented in existing CCD manufacturing facilities at low cost.
CCDs with sub-electron readout noise have a broad range of applications including particle physics (\eg, ultra-low-noise searches for dark matter and neutrinos) and astronomy (\eg, direct imaging and spectroscopy of exoplanets).

\begin{table}
\centering
\caption{\label{tab:skipper}
Skipper CCD Detector Characteristics
}
\begin{tabular}{l c c c}
\hline
Characteristic  & Value  & Unit\\
\hline \hline
Format & $4126 \times 866$ & pixels\\
Pixel Scale & 15 & $\um$ \\
Thickness & $200$ & $\um$\\
Operating Temperature & 140 & Kelvin \\
Number of Amplifiers & 4 & \\
Dark Current\footnotemark[1] & $< 10^{-3}$ & $\e/\pix/{\rm day}$ \\
Readout Time (1 sample) & 10 & $\us$/pix/amp \\
Readout Noise (1 sample) & $3.55 $ & $\ermspix$ \\
Readout Noise (4000 samples) & $0.068$ & $\ermspix$ \\
\hline
\end{tabular}
\footnotetext[1]{The upper limit on dark current comes from measurements on a similar CCD used by the DAMIC experiment \cite{Aguilar-Arevalo:2016zop}. 
}
\end{table}

\section{Technical Description}

\subsection{CCD detector}

The detector studied here is a p-channel CCD fabricated on high resistivity ($\sim$10~k$\Omega$~cm) n-type silicon that was fully depleted at a substrate voltage of $40$~V.  
The sensor is 200~$\mu$m thick and composed of $15 \um \times 15 \um$ square pixels arranged in a $4126 \times 866$ array. 
The characteristics of the Skipper CCD are collected in Table~\ref{tab:skipper}.  
To reduce the number of electrons promoted to the silicon conduction band by thermal fluctuations (``dark current''), the sensor is operated at low temperatures. 
Here we operate the sensor at 140~K, but this could be lowered to  $\sim 100$~K before charge-transfer efficiency is significantly reduced.  
As we discuss further below, the dark current may be the limiting factor for some applications though significant investment has been made to minimize it \citep{Holland:1989,Holland:2003}. 

\begin{figure}[!b]
\centering
\includegraphics[width=\columnwidth]{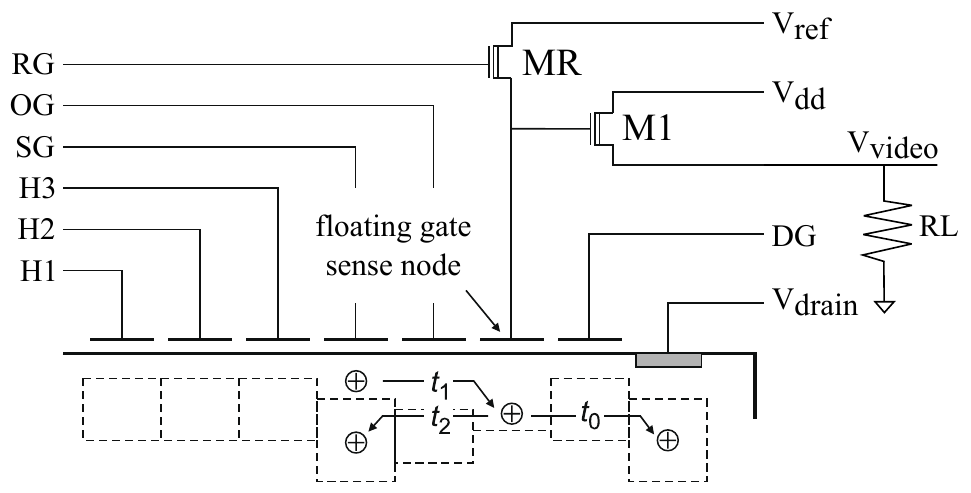}
\caption{Simplified diagram of the Skipper CCD output stage. H1, H2 and H3 are the horizontal register clock phases. MR is a switch to reset the sense node to $V_{\rm ref}$. M1 is a MOSFET in a source follower configuration. Due to its floating gate, the Skipper readout performs a non-destructive measurement of the charge at the SN.}
\label{fig:outStage}
\end{figure}

Figure \ref{fig:outStage} shows a simplified diagram of the Skipper CCD output stage. 
At $t_0$,  all the charge is drained from the sense node (SN) to $V_{\rm drain}$ by applying a pulse to the dump gate (DG), and the SN voltage is restored to $V_{\rm ref}$ with a pulse to the reset gate (RG). 
At $t_{1}$, the summing-well gate (SG) phase is raised to transfer the charge packet to the SN and conclude the readout of the first sample. 
To take the second sample, the output gate (OG) and SG phase are lowered at $t_{2}$, moving the charge packet in the SN back under the SG phase and the reference voltage of the SN is restored applying a pulse to the RG. This cycle can be repeated to sample the same charge packet multiple times. A more detailed description of the Skipper output stage can be found in~\cite{Moroni:2012}.

The CCD is divided into four rectangular regions of 2063$\times$433 pixels, each of which is read by an independent amplifier possessing a distinct readout design.
The most important difference between the readout designs tested is the size of the floating gate. 
Smaller floating gates have smaller capacitance and higher gain, but can be subject to charge transfer inefficiency and reduced full-well capacity. 
However, none of the designs tested showed any adverse effects from reducing the size of the readout structures. 
The results discussed here were collected from the readout stage with the smallest floating gate, with an area of $15 \um \times 4 \um$.
Additional improvements in the gain are expected from further reduction of the size of the floating gate, which will be explored in future generations of Skipper CCDs.

The output amplifiers of our chip have an impedance of $\roughly 2$~k$\Omega$ and can only drive a short cable of a few cm without significant signal degradation. 
To reduce the load on these amplifiers, junction gate field-effect transistor (JFET) source followers that operate at low temperatures ($\sim$120~K) were placed next to each analog output. 
The JFETs lower the output impedance making it possible to drive signals over a 50~cm flex circuit and reach precision operational amplifiers at room temperature. 
These operational amplifiers have a noise density of 1.1~nV/$\sqrt{\textrm{Hz}}$ and are required to reduce the impact of internal sources of noise in the readout electronics.

\subsection{Readout electronics}

The readout electronics are based on the Monsoon system developed for the Dark Energy Camera~\cite{Mclean:2012pka,Flaugher:2015}. 
This system can be adapted to independently control the extra gates in the Skipper CCD and provides simultaneous digitization of the four amplifier channels of our sensor. 
To fully benefit from the multiple sampling capabilities of the Skipper CCD, the noise for each sample must be uncorrelated with the noise of the other samples. 
We found that the electronics initially showed high levels of correlated noise between samples. 
This problem was solved by replacing all switching power supplies with low-noise laboratory power supplies that have ripple noise V$_\textrm{rms}<$350~$\mu$V  and peak-to-peak ripple voltage V$_\textrm{pp}<2$~mV \citep{Haro:2016}.

\subsection{Signal processing and performance}

\begin{figure}
\centering
\includegraphics[width=0.5\textwidth]{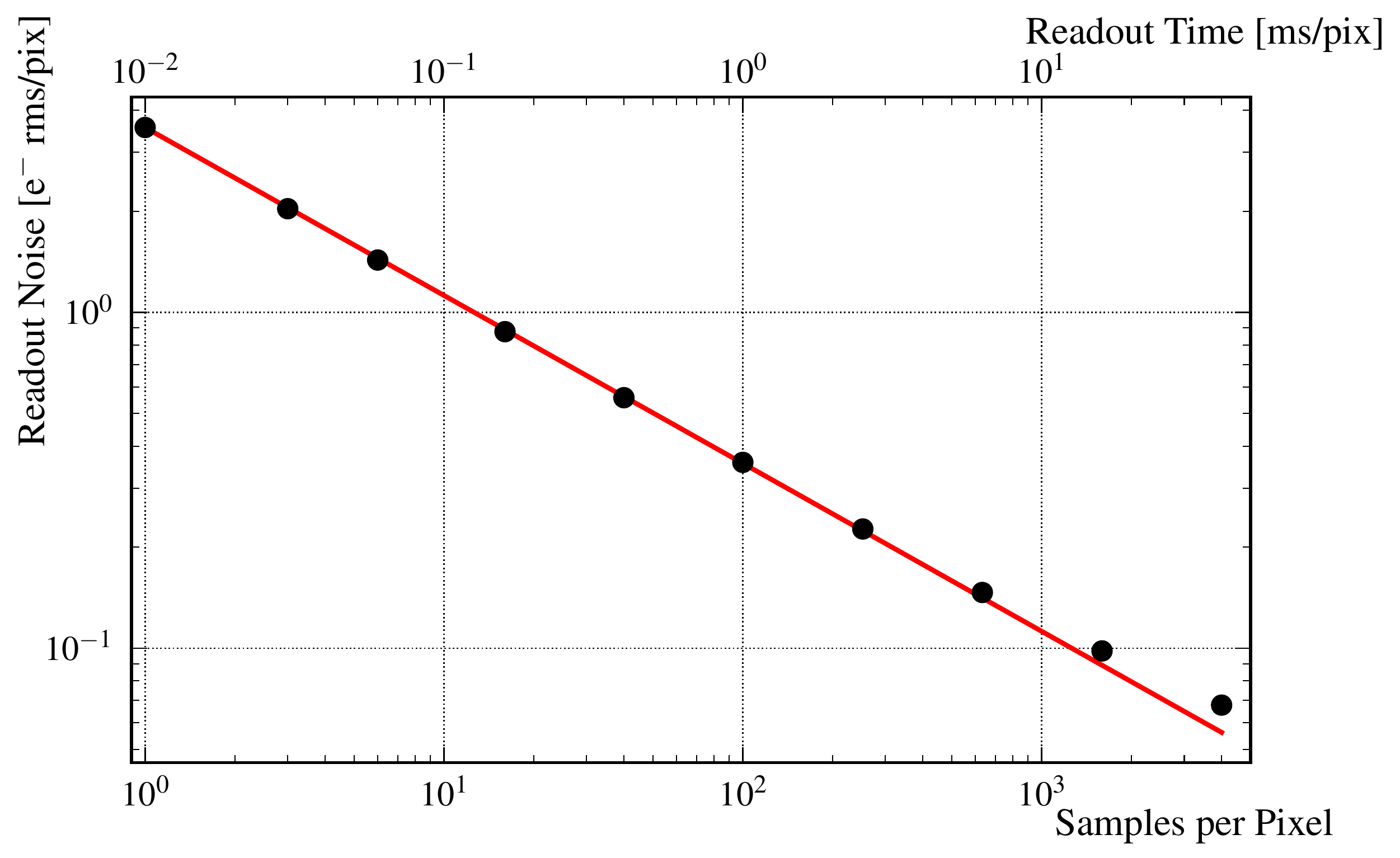}
\caption{
Measured readout noise as a function of the number of non-destructive readout samples per pixel for the Skipper CCD. Black points show the standard deviation of the empty pixels distribution as a function of the number of averaged samples. The red line is the theoretical expectation assuming independent, uncorrelated samples (Eq.~\eqref{eq:noise}). The number of readout samples can be dynamically configured on a per pixel basis.
}
\label{fig:noiseVsN}
\end{figure}

One major advantage of the non-destructive readout technique is that individual samples should be uncorrelated measurements of the charge in each pixel.
For uncorrelated Gaussian readout noise, the optimal estimator for the measured pixel value is the average value of the samples taken. 
In this case, the standard deviation, $\sigma$, of the effective readout noise distribution after averaging $N$ samples per pixel is simply
\begin{equation}
\label{eq:noise}
\sigma=\frac{\sigma_{1}}{\sqrt{N}},
\end{equation}
where $\sigma_{1}$ is the standard deviation of the readout noise for a single sample of the pixel. 
Figure~\ref{fig:noiseVsN} shows that the measured performance of the Skipper CCD closely follows the prediction for uncorrelated Gaussian noise, deviating only slightly for $N \gtrsim 2000$ samples. 
The agreement between the theoretical prediction and the measured performance indicates that the assumption of uncorrelated Gaussian noise is valid, and more sophisticated signal processing techniques are not necessary.

Another advantage of non-destructive readout is that it is possible to dynamically configure the number of samples taken per pixel.
This allows the readout noise to be adjusted on a per pixel basis.
The simplest application of this procedure is to repeatedly sample a predetermined subset of pixels where low readout noise is desired, while reading out the rest of the detector quickly with fewer samples per pixel.\footnote{This approach of ``skipping'' a large number of pixels is the origin for the name ``Skipper'' CCD~\citep{Janesick:2001}.}
The flexibility of the non-destructive readout technique also allows the design of sampling schemes that are based on the value of a pixel. 
For example, the measured value from the first readout of a pixel can be used to determine the number of subsequent readout samples to be taken.
In this way, the readout noise can be adjusted based on the amount of charge in a pixel.
Dynamic sampling schemes can greatly reduce the readout time for the Skipper CCD when sub-electron noise is not required in all pixels simultaneously. 

\subsection{Readout Time}
The Skipper CCD studied here has a single-sample readout noise of $\sigma_1 = 3.55\ermspix$ with a readout time of $10\us/\pix/\spl$. 
To reach a readout noise of $\sigma < 0.1 \ermspix$ requires $\roughly 1200$ samples per pixel, corresponding to a readout time of $12\ms/\pix$.
The wide-format detector described here can be read out in 3 hours with 4 amplifiers.
The readout time scales linearly with the number of samples and inversely with the square of the readout noise. 
The configurability of the Skipper readout allows the number of samples and readout time to be dynamically adjusted for applications where a larger readout noise is sufficient.
In addition, there are several mechanisms that can be used to further reduce the readout time:
\begin{enumerate}
\item The single-sample readout noise and readout time can be decreased by using lower capacitance amplifiers with higher gain. Current CCDs using these amplifiers routinely reach $2\e$ rms/pix with a readout time of 4\us per pixel per sample~\cite{Bebek:2017}. 
A Skipper CCD with this performance would be able to reach $\sigma < 0.1 \ermspix$ 7.5 times faster than the device studied here.
\item The number of readout channels and amplifiers can be increased. Commercial devices are available with 16 amplifiers~\citep{LSST:2015}, and sensors with $>48$ on-chip amplifiers are well within the reach of existing technology \citep{Doering:2011}.
The readout time will decrease linearly with the number of amplifiers. 
\end{enumerate}
 
An $\mathcal{O}(100)$ decrease in the Skipper CCD readout time can be envisioned by combining these two techniques.
In addition, for specific applications a frame shifting readout would allow readout to be performed concurrently with the next exposure.
Finally, as discussed in the previous section, the flexibility of the Skipper readout can greatly reduce the readout time for applications where only a fraction of the pixels require ultra-low readout noise.

\subsection{Impact of dark current}

For a detector with sub-electron readout noise, the dark current can become the dominant source of background in rare event searches.
A Skipper CCD with readout noise $\sigma \lesssim 0.1 \ermspix$ will misclassify a pixel with $n$ electrons as having $n+1$ electrons with a probability of $p\sim3\times10^{-7}$ ($5\sigma$).
This probability can be further reduced by only keeping pixels with measured values that are less than 3$\sigma$ away from an integer charge (i.e., only keeping pixels with measured values between $n-0.3$ and $n + 0.3$, where $n$ is any non-negative integer).
By applying this quality cut, the misclassification probability is reduced to $p\sim10^{-12}$ ($\roughly 7\sigma$), while the efficiency is kept above 99.7\%.
Under these conditions the sensor operates in the single-electron counting regime (\ie, it is a zero-noise detector), and readout noise no longer limits the energy/charge threshold of the sensor.
In particular, the detector is sensitive to all particle interactions with energy deposition above the silicon band gap of $1.1\eV$. 

The number of dark-current electrons that accumulate in a pixel is expected to follow a Poisson distribution with a mean equal to the dark-current rate multiplied by the exposure time.
The minimum exposure time for a zero-noise Skipper CCD is set by the readout time. 
The Skipper CCD described here can be read out in 3 hours with a readout noise of $\sigma = 0.1 \ermspix$.
Current experiments using fully depleted scientific CCDs similar to the Skipper CCD studied here have measured an upper limit on the dark-current rate of $\lesssim 10^{-3} \epixday$~\cite{Aguilar-Arevalo:2016zop}, with no evidence for deviations from a Poisson distribution.
Assuming the counts from the dark current follow a Poisson distribution, the upper limit on the expected number of pixels with an accumulated charge of $\geq 1\e$ ($\geq 2\e$) is $\roughly 450$ ($\roughly 0.03$) per exposure. 
However, the theoretically expected dark-current rate for a CCD operating at 120~K is $\roughly 10^{-7}$~\cite{Janesick:2001}, which would predict 0.04~pixels with $\geq 1\e$ and $O(10^{-10})$ pixels with $\geq 2\e$ per exposure. 

Depending on the application and the magnitude of the dark current, a faster readout time with larger noise may be preferred.
For dark matter searches (see below) the most important parameter is the charge or energy threshold, which should be set as low as possible to capture as many dark matter events as possible~\cite{Essig:2015cda}. 
The Skipper CCD tested here can be read in 1.5-hours with a noise of $0.14\ermspix$. 
At this noise level, a selection cut is required to avoid misclassification of the events in the right-tail of the lowest charge bin dominated by dark current events (leakage from the 1\e or 2\e bin, depending on the dark current). 
By requiring the measured charge be $>0.84\e$ ($> 6\sigma$) above the dark-current bin, a rejection of $O(10^{9})$ is obtained while keeping a selection efficiency of 87\% on the first significant bin (2\e or 3\e) for rare events searches.

To demonstrate the power of Skipper CCDs for rare event searches, consider a $100$\,g detector consisting of 80 3.5-Mpix Skipper CCDs, which are continuously read out for a total accumulated exposure time of 1~year. 
Table~\ref{tab:dc} shows the number of Skipper CCD pixels expected to exceed a given charge threshold (\ie, 1\e, 2\e, or 3\e) for three different values of the dark current.
The Skipper readout allows a threshold of 3\e to be achieved for a dark-current value comparable to the upper limit reported in similar CCD detectors~\cite{Aguilar-Arevalo:2016zop}, while a dark current of $\lesssim 10^{-5} \epixday$ would reduce the expected number of pixels with $\geq 2\e$ from dark current to a negligible level.
The number of dark-current events will decrease linearly with the readout time, assuming the total integrated exposure is fixed.

\begin{table}
\centering
\begin{tabular}{c c l l}
\hline
Dark Current & $\geq 1\e$ & $\geq 2\e$ & $\geq 3\e$ \\
(\epixday) & (pix) & (pix) & (pix) \\
\hline \hline
$10^{-3}$ & ~$1\times10^{8}$ & ~$3\times10^{3}$ & ~$7\times10^{-2}$ \\
$10^{-5}$ & ~$1\times10^{6}$ & ~$3\times10^{-1}$  & ~$7\times10^{-8}$ \\
$10^{-7}$ & ~$1\times10^{4}$ & ~$3\times10^{-5}$& ~$7\times10^{-14}$ \\
\hline
\end{tabular}
\caption{
\label{tab:dc}
Number of pixels expected to exceed a given charge threshold as a function of the CCD dark current. 
We assume a benchmark experiment with a total exposure of 0.1~kg-years.  
This could be built using 80 Skipper CCDs with the same readout design and format of the detector presented here ($4126 \times 866$ pixels), but with a thickness of $650\um$. 
The aggregated total of 280 Mpix could be read in 1.5 hours with a readout noise of $0.14 \ermspix$.
A dark current of $10^{-3} \epixday$ corresponds to the upper limit reported by DAMIC~\cite{Aguilar-Arevalo:2016zop}. 
A dark current of $10^{-5} \epixday$ would allow a threshold of $2 \e$ in rare events searches, while the theoretical expectation for the minimum dark-current rate is $10^{-7} \epixday$.  
}
\end{table}

\section{Applications}

\subsection{Particle Physics}

The ability to precisely count the number of electrons in a pixel with a threshold as low as 2\e or 3\e  would have a significant and immediate impact on rare event searches. 
In particular, the Skipper CCD allows for the construction of a new dark matter direct detection experiment with unprecedented sensitivity to several classes of particle dark matter.  
This includes dark matter particles with masses $\gtrsim 1 \MeV$ that scatter off an  electron~\cite{Essig:2011nj,Graham:2012su,Lee:2015qva,Essig:2015cda}, and bosonic dark matter particles (pseudoscalar, scalar, or vector particles) with masses greater than a few eV that get absorbed by an electron~\cite{An:2013yua,An:2014twa,Bloch:2016sjj,Hochberg:2016sqx}. 
Dark matter that scatters elastically off nuclei or that scatters inelastically 
off nuclei while emitting a photon~\cite{Kouvaris:2016afs}, can produce a measurable 
signal in a Skipper CCD for dark matter masses as low as a few hundred MeV  or 
a few tens of MeV, respectively.  

Sub-GeV dark matter can scatter off an electron in the valence band of silicon,
promoting it to the conduction band. 
The typical recoil energy of the scattered electron is a few eV, with the dark-matter-electron scattering rate falling steeply for larger recoil energies. A threshold of 2\e (3\e), which corresponds to an electron recoil energy of about 4.7~eV (8.3~eV), would thus capture the majority of dark matter-electron scattering events for dark matter masses above $\mathcal{O}$(MeV)~\cite{Essig:2015cda} (this assumes that 3.6~eV of energy above the silicon band gap is needed to create each additional electron-hole pair~\cite{Klein:1968,ExptGaps}).   
The readout noise of previous silicon CCD detectors required a higher threshold of $\sim 11$\e (40~eV)~\cite{Barreto:2011zu}.
These detectors thus had a relatively high mass-threshold and limited cross-section sensitivity to dark matter-electron scattering~\cite{Essig:2015cda}.  
A 2\e or 3\e threshold in silicon is also lower than previously achieved using xenon time projection chambers (TPCs), whose threshold is set by the minimum ionization energy of xenon of 12.1 eV. 
Moreover, while existing XENON10 data~\cite{Angle:2011th} already constrain dark matter as light as $\sim 5\ $MeV~\cite{Essig:2012yx,Essig:2017kqs}, a detector-specific background limited the sensitivity and discovery potential of this experiment (further research is needed to determine whether this spurious background can be reduced in xenon TPCs~\cite{Sorensen:2017ymt}). 
In contrast, a Skipper CCD  
would have significant discovery potential and is expected to have no backgrounds for exposures of $\sim 100\ $gram-years or less, assuming similar radiogenic background levels as achieved in other experiments~\cite{Agnese:2015nto,Aalseth:2012if,Abgrall:2016tnn,Armengaud:2016aoz}. 
Such a detector can thus explore orders of magnitude of new dark matter parameter space~\cite{Essig:2015cda}.
 
Bosonic dark matter with masses as low as the silicon band gap could also be absorbed by an electron that is subsequently excited to the conduction band. 
In this inelastic process, the excitation energy is equal to the mass of the 
dark matter, and the lowest detectable mass is determined by the experimental threshold. The current best limits on dark-photon dark matter are set by the DAMIC experiment, which constrains masses down to 3 eV~\cite{Aguilar-Arevalo:2016zop}. However, this 
measurement was dominated by the readout noise and had no discovery potential.  
A Skipper CCD would dramatically improve the sensitivity to such 
low-mass dark matter and allow for a discovery.  

If dark matter scatters elastically off a nucleus, the resulting nuclear recoil can 
produce a measurable ionization signal for dark matter masses as low as a few hundred MeV. 
The precise mass threshold depends on the ionization efficiency, which is not accurately known at low energies.  The dark matter can also scatter inelastically off 
a nucleus with the nucleus emitting a photon in the process.  Depending on the 
energy of the photon, some number of electrons would be produced in a pixel. 
Further study is required, but a mass threshold below 100~MeV for detecting such an 
interaction is likely. 

In addition to dark matter searches, CCD sensors have found an exciting new application as a target material for coherent neutrino-nucleus scattering~\cite{Moroni:2014wia, DAMIC:2013,Aguilar-Arevalo:2016qen}. These experiments have a low event rate and do not require fast readout times.  The Skipper CCDs would provide a lower energy threshold, allowing for the exploration of new neutrino parameter space and non-standard neutrino interactions~\cite{Harnik:2012ni}.

\subsection{Astronomy}

Silicon CCDs have ushered in an era of precision astronomy and cosmology.
To assess the impact of Skipper CCDs for astronomical applications, we examine how the signal-to-noise ratio (SNR) per pixel depends on the readout noise. Specifically, we define
\begin{equation}
{\rm SNR} = \frac{r_{\rm sig} t_{\rm exp}}{\sigma_{\rm tot}}
\end{equation}
\noindent where $r_{\rm sig}$ is the count rate of the signal source, $t_{\rm exp}$ is the exposure time, and $\sigma_{\rm tot}$ is the total noise contribution per pixel, 
\begin{equation}
\sigma_{\rm tot}^2 =  (r_{\rm sig} + r_{\rm bkg})t_{\rm exp} + r_{\rm dark} t_{\rm exp} + \sigma_{\rm read}^2\,.
\end{equation}
The first term in this equation corresponds to Poisson shot noise from the signal source and sky background (including other sources of incident light).
The second term is a contribution from the dark current (generally sub-dominant for astronomical applications) and the last term is the readout noise per pixel, $\sigma_{\rm read}$.
In order to reach a given SNR threshold requires an exposure time expressed as,
\begin{align}
t_{\rm exp} =& \frac{{\rm SNR}^2 r}{2 r_{\rm sig}^2} \left (1
 + \sqrt{1 + \frac{4 r_{\rm sig}^2 \sigma_{\rm read}^2}{r^2 {\rm SNR}^2} } \right )\,.
\end{align}
\noindent where we have expressed the total counts rate, $r = (r_{\rm sig} + r_{\rm bkg} + r_{\rm dark})$. 

By using the non-destructive readout of the Skipper CCD, it is possible to reduce the readout noise from $\sigma_{\rm read} = \sigma_1$ to $\sigma_{\rm read} = \sigma \ll 1$ (Eq. \ref{eq:noise}). 
A drastic reduction in readout noise can significantly reduce the exposure time required to reach a given SNR if,
\begin{equation}
\sigma_{1} > \frac{ {\rm SNR}\, r}{2r_{\rm sig}} \, .
\end{equation}
\noindent Assuming a single-sample readout noise of $\sigma_{1} = 3.55 \ermspix$, the applications that will benefit most from reduced readout noise will be signal dominated ($r \approx r_{\rm sig}$), but possess a low signal to noise ratio (${\rm SNR} \lesssim 7$). 

One exciting science case that will operate in the very low SNR regime is space-based imaging and spectroscopy of terrestrial exoplanets in the habitable zones of nearby stars~\cite{DecadalReview2010}. 
The photon flux from exo-Earths is expected to be of order 1 per several minutes, necessitating the use of ultra-low noise detectors \cite{ExoPlanet:2017}.
A detector with sub-electron readout noise could reduce exposure times by a factor of two~\cite{Shaklan:2013}.
Skipper CCDs are easily manufactured with large formats (\ie, 2k$\times$2k pix) and are stable over a large dynamic range.
In addition, thick fully depleted CCDs can achieve high quantum efficiency between 0.87\um and 1\um where several important spectral lines from water reside \cite{ExoPlanet:2017}.

Another potential application for ultra-low-noise detectors is in the study of short duration variable sources.
Limited exposure times naturally produce signal-dominated, low SNR observations where the impact of readout noise is maximized.
However, the application of Skipper CCDs to short duration temporal sampling would require a significant decrease in readout time, which may be achieved through the techniques described above (e.g., reductions in single-sample noise, multiple amplifiers, selective readout, etc.) 
A factor of 100 decrease in readout time for the detector discussed here would make Skipper CCDs a competitive technology for time-domain astronomy.

\section{Conclusions}
We have demonstrated single electron and single photon counting over a large dynamic range in a thick, fully depleted, silicon Skipper CCD.
Using a non-destructive readout technique, we have achieved a readout noise of $0.068 \ermspix$, far in the sub-electron regime.
This novel detector has immediate applications in rare event searches for {\emph{e.g.}} dark matter and neutrinos, but may also prove useful for astronomy and other scientific applications.

\section*{Acknowledgments}
We wish to recognize the joy and passion of Yann Guardincerri. 
He will be dearly missed as a friend and a colleague.
The authors gratefully acknowledge Juan Estrada for many useful conversations, and 
Chris Bebek for fundamental contributions to the development of the Skipper CCD.
The authors thank Ting Li for helpful discussions on astronomical applications.
R.E.\ is supported by the DOE Early Career research program DESC0008061 and through a Sloan Foundation Research Fellowship.
 T.V.\ is supported by the I-CORE Program of
the Planning Budgeting Committee and the Israel Science
Foundation (grant No. 1937/12), the European Research Council
(ERC) under the EU Horizon 2020 Programme (ERCCoG-2015-Proposal n.682676 LDMThExp), and the German-Israeli Foundation (grant No.\ I-1283-303.7/2014).
CCD development was supported by the Lawrence Berkeley National Lab Director, Office of Science, of the U.S. Department of Energy under
Contract No.\ DE-AC02-05CH11231. 
This work was supported by Fermilab LDRD under DOE Contract No.\ DE-AC02-07CH11359. 

\bibliography{main}

\begin{thebibliography}{47}%
\makeatletter
\providecommand \@ifxundefined [1]{%
 \@ifx{#1\undefined}
}%
\providecommand \@ifnum [1]{%
 \ifnum #1\expandafter \@firstoftwo
 \else \expandafter \@secondoftwo
 \fi
}%
\providecommand \@ifx [1]{%
 \ifx #1\expandafter \@firstoftwo
 \else \expandafter \@secondoftwo
 \fi
}%
\providecommand \natexlab [1]{#1}%
\providecommand \enquote  [1]{``#1''}%
\providecommand \bibnamefont  [1]{#1}%
\providecommand \bibfnamefont [1]{#1}%
\providecommand \citenamefont [1]{#1}%
\providecommand \href@noop [0]{\@secondoftwo}%
\providecommand \href [0]{\begingroup \@sanitize@url \@href}%
\providecommand \@href[1]{\@@startlink{#1}\@@href}%
\providecommand \@@href[1]{\endgroup#1\@@endlink}%
\providecommand \@sanitize@url [0]{\catcode `\\12\catcode `\$12\catcode
  `\&12\catcode `\#12\catcode `\^12\catcode `\_12\catcode `\%12\relax}%
\providecommand \@@startlink[1]{}%
\providecommand \@@endlink[0]{}%
\providecommand \url  [0]{\begingroup\@sanitize@url \@url }%
\providecommand \@url [1]{\endgroup\@href {#1}{\urlprefix }}%
\providecommand \urlprefix  [0]{URL }%
\providecommand \Eprint [0]{\href }%
\providecommand \doibase [0]{http://dx.doi.org/}%
\providecommand \selectlanguage [0]{\@gobble}%
\providecommand \bibinfo  [0]{\@secondoftwo}%
\providecommand \bibfield  [0]{\@secondoftwo}%
\providecommand \translation [1]{[#1]}%
\providecommand \BibitemOpen [0]{}%
\providecommand \bibitemStop [0]{}%
\providecommand \bibitemNoStop [0]{.\EOS\space}%
\providecommand \EOS [0]{\spacefactor3000\relax}%
\providecommand \BibitemShut  [1]{\csname bibitem#1\endcsname}%
\let\auto@bib@innerbib\@empty
\bibitem [{\citenamefont {Boyle}\ and\ \citenamefont
  {Smith}(1970)}]{Boyle:1970}%
  \BibitemOpen
  \bibfield  {author} {\bibinfo {author} {\bibfnamefont {W.~S.}\ \bibnamefont
  {Boyle}}\ and\ \bibinfo {author} {\bibfnamefont {G.~E.}\ \bibnamefont
  {Smith}},\ }\href {\doibase 10.1002/j.1538-7305.1970.tb01790.x} {\bibfield
  {journal} {\bibinfo  {journal} {The Bell System Technical Journal}\ }\textbf
  {\bibinfo {volume} {49}},\ \bibinfo {pages} {587} (\bibinfo {year}
  {1970})}\BibitemShut {NoStop}%
\bibitem [{\citenamefont {{{Amelio}, G.~F. and {Tompsett}, M.~F. and {Smith},
  G.~E.}}(1970)}]{Amelio:1970}%
  \BibitemOpen
  \bibfield  {author} {\bibinfo {author} {\bibnamefont {{{Amelio}, G.~F. and
  {Tompsett}, M.~F. and {Smith}, G.~E.}}},\ }\href@noop {} {\bibfield
  {journal} {\bibinfo  {journal} {Bell Syst. Tech. J.}\ }\textbf {\bibinfo
  {volume} {49}},\ \bibinfo {pages} {593} (\bibinfo {year} {1970})}\BibitemShut
  {NoStop}%
\bibitem [{\citenamefont {Damerell}\ \emph {et~al.}(1981)\citenamefont
  {Damerell}, \citenamefont {Farley}, \citenamefont {Gillman},\ and\
  \citenamefont {Wickens}}]{damerell:1981}%
  \BibitemOpen
  \bibfield  {author} {\bibinfo {author} {\bibfnamefont {C.}~\bibnamefont
  {Damerell}}, \bibinfo {author} {\bibfnamefont {F.}~\bibnamefont {Farley}},
  \bibinfo {author} {\bibfnamefont {A.}~\bibnamefont {Gillman}}, \ and\
  \bibinfo {author} {\bibfnamefont {F.}~\bibnamefont {Wickens}},\ }\href@noop
  {} {\bibfield  {journal} {\bibinfo  {journal} {Nuclear Instruments and
  Methods in Physics Research}\ }\textbf {\bibinfo {volume} {185}},\ \bibinfo
  {pages} {33} (\bibinfo {year} {1981})}\BibitemShut {NoStop}%
\bibitem [{\citenamefont {{{Janesick}, J.~R.}}(2001)}]{Janesick:2001}%
  \BibitemOpen
  \bibfield  {author} {\bibinfo {author} {\bibnamefont {{{Janesick}, J.~R.}}},\
  }\href@noop {} {\emph {\bibinfo {title} {{Scientific Charge Coupled
  Devices}}}}\ (\bibinfo  {publisher} {SPIE Publications},\ \bibinfo {year}
  {2001})\BibitemShut {NoStop}%
\bibitem [{\citenamefont {White}\ \emph {et~al.}(1974)\citenamefont {White},
  \citenamefont {Lampe}, \citenamefont {Blaha},\ and\ \citenamefont
  {Mack}}]{CDS:1050448}%
  \BibitemOpen
  \bibfield  {author} {\bibinfo {author} {\bibfnamefont {M.~H.}\ \bibnamefont
  {White}}, \bibinfo {author} {\bibfnamefont {D.~R.}\ \bibnamefont {Lampe}},
  \bibinfo {author} {\bibfnamefont {F.~C.}\ \bibnamefont {Blaha}}, \ and\
  \bibinfo {author} {\bibfnamefont {I.~A.}\ \bibnamefont {Mack}},\ }\href
  {\doibase 10.1109/JSSC.1974.1050448} {\bibfield  {journal} {\bibinfo
  {journal} {IEEE Journal of Solid-State Circuits}\ }\textbf {\bibinfo {volume}
  {9}},\ \bibinfo {pages} {1} (\bibinfo {year} {1974})}\BibitemShut {NoStop}%
\bibitem [{\citenamefont {Janesick}\ and\ \citenamefont
  {Tower}(2016)}]{Janesick:2016}%
  \BibitemOpen
  \bibfield  {author} {\bibinfo {author} {\bibfnamefont {J.}~\bibnamefont
  {Janesick}}\ and\ \bibinfo {author} {\bibfnamefont {J.}~\bibnamefont
  {Tower}},\ }\href {\doibase 10.3390/s16050688} {\bibfield  {journal}
  {\bibinfo  {journal} {Sensors}\ }\textbf {\bibinfo {volume} {16}},\ \bibinfo
  {pages} {688} (\bibinfo {year} {2016})}\BibitemShut {NoStop}%
\bibitem [{\citenamefont {Bebek}\ \emph {et~al.}(2017)\citenamefont {Bebek},
  \citenamefont {Emes}, \citenamefont {Groom}, \citenamefont {Haque},
  \citenamefont {Holland}, \citenamefont {Jelinsky}, \citenamefont {Karcher},
  \citenamefont {Kolbe}, \citenamefont {Lee}, \citenamefont {Palaio},
  \citenamefont {Schlegel}, \citenamefont {Wang}, \citenamefont {Groulx},
  \citenamefont {Frost}, \citenamefont {Estrada},\ and\ \citenamefont
  {Bonati}}]{Bebek:2017}%
  \BibitemOpen
  \bibfield  {author} {\bibinfo {author} {\bibfnamefont {C.}~\bibnamefont
  {Bebek}}, \bibinfo {author} {\bibfnamefont {J.}~\bibnamefont {Emes}},
  \bibinfo {author} {\bibfnamefont {D.}~\bibnamefont {Groom}}, \bibinfo
  {author} {\bibfnamefont {S.}~\bibnamefont {Haque}}, \bibinfo {author}
  {\bibfnamefont {S.}~\bibnamefont {Holland}}, \bibinfo {author} {\bibfnamefont
  {P.}~\bibnamefont {Jelinsky}}, \bibinfo {author} {\bibfnamefont
  {A.}~\bibnamefont {Karcher}}, \bibinfo {author} {\bibfnamefont
  {W.}~\bibnamefont {Kolbe}}, \bibinfo {author} {\bibfnamefont
  {J.}~\bibnamefont {Lee}}, \bibinfo {author} {\bibfnamefont {N.}~\bibnamefont
  {Palaio}}, \bibinfo {author} {\bibfnamefont {D.}~\bibnamefont {Schlegel}},
  \bibinfo {author} {\bibfnamefont {G.}~\bibnamefont {Wang}}, \bibinfo {author}
  {\bibfnamefont {R.}~\bibnamefont {Groulx}}, \bibinfo {author} {\bibfnamefont
  {R.}~\bibnamefont {Frost}}, \bibinfo {author} {\bibfnamefont
  {J.}~\bibnamefont {Estrada}}, \ and\ \bibinfo {author} {\bibfnamefont
  {M.}~\bibnamefont {Bonati}},\ }\href
  {http://stacks.iop.org/1748-0221/12/i=04/a=C04018} {\bibfield  {journal}
  {\bibinfo  {journal} {Journal of Instrumentation}\ }\textbf {\bibinfo
  {volume} {12}},\ \bibinfo {pages} {C04018} (\bibinfo {year}
  {2017})}\BibitemShut {NoStop}%
\bibitem [{\citenamefont {{Janesick}}\ \emph {et~al.}(1990)\citenamefont
  {{Janesick}}, \citenamefont {{Elliott}}, \citenamefont {{Dingizian}},
  \citenamefont {{Bredthauer}},\ and\ \citenamefont
  {{Chandler}}}]{Janesick:1990}%
  \BibitemOpen
  \bibfield  {author} {\bibinfo {author} {\bibfnamefont {J.~R.}\ \bibnamefont
  {{Janesick}}}, \bibinfo {author} {\bibfnamefont {T.}~\bibnamefont
  {{Elliott}}}, \bibinfo {author} {\bibfnamefont {A.}~\bibnamefont
  {{Dingizian}}}, \bibinfo {author} {\bibfnamefont {R.~A.}\ \bibnamefont
  {{Bredthauer}}}, \ and\ \bibinfo {author} {\bibfnamefont {C.~E.}\
  \bibnamefont {{Chandler}}},\ }\href@noop {} {\bibfield  {journal} {\bibinfo
  {journal} {SPIE}\ }\textbf {\bibinfo {volume} {1242}},\ \bibinfo {pages}
  {223} (\bibinfo {year} {1990})}\BibitemShut {NoStop}%
\bibitem [{\citenamefont {Wen}(1974)}]{Wen:1974}%
  \BibitemOpen
  \bibfield  {author} {\bibinfo {author} {\bibfnamefont {D.}~\bibnamefont
  {Wen}},\ }\href {\doibase 10.1109/JSSC.1974.1050535} {\bibfield  {journal}
  {\bibinfo  {journal} {IEEE Journal of Solid-State Circuits}\ }\textbf
  {\bibinfo {volume} {9}},\ \bibinfo {pages} {410} (\bibinfo {year}
  {1974})}\BibitemShut {NoStop}%
\bibitem [{\citenamefont {{Chandler}}\ \emph {et~al.}(1990)\citenamefont
  {{Chandler}}, \citenamefont {{Bredthauer}}, \citenamefont {{Janesick}},
  \citenamefont {Westphal},\ and\ \citenamefont {{Gunn}}}]{Chandler:1990}%
  \BibitemOpen
  \bibfield  {author} {\bibinfo {author} {\bibfnamefont {C.~E.}\ \bibnamefont
  {{Chandler}}}, \bibinfo {author} {\bibfnamefont {R.~A.}\ \bibnamefont
  {{Bredthauer}}}, \bibinfo {author} {\bibfnamefont {J.~R.}\ \bibnamefont
  {{Janesick}}}, \bibinfo {author} {\bibfnamefont {J.~A.}\ \bibnamefont
  {Westphal}}, \ and\ \bibinfo {author} {\bibfnamefont {J.~E.}\ \bibnamefont
  {{Gunn}}},\ }\href@noop {} {\bibfield  {journal} {\bibinfo  {journal} {SPIE}\
  }\textbf {\bibinfo {volume} {1242}},\ \bibinfo {pages} {238} (\bibinfo {year}
  {1990})}\BibitemShut {NoStop}%
\bibitem [{\citenamefont {Fernandez~Moroni}\ \emph {et~al.}(2012)\citenamefont
  {Fernandez~Moroni}, \citenamefont {Estrada}, \citenamefont {Cancelo},
  \citenamefont {Holland}, \citenamefont {Paolini},\ and\ \citenamefont
  {Diehl}}]{Moroni:2012}%
  \BibitemOpen
  \bibfield  {author} {\bibinfo {author} {\bibfnamefont {G.}~\bibnamefont
  {Fernandez~Moroni}}, \bibinfo {author} {\bibfnamefont {J.}~\bibnamefont
  {Estrada}}, \bibinfo {author} {\bibfnamefont {G.}~\bibnamefont {Cancelo}},
  \bibinfo {author} {\bibfnamefont {S.~E.}\ \bibnamefont {Holland}}, \bibinfo
  {author} {\bibfnamefont {E.~E.}\ \bibnamefont {Paolini}}, \ and\ \bibinfo
  {author} {\bibfnamefont {H.~T.}\ \bibnamefont {Diehl}},\ }\href {\doibase
  10.1007/s10686-012-9298-x} {\bibfield  {journal} {\bibinfo  {journal} {Exper.
  Astron.}\ }\textbf {\bibinfo {volume} {34}},\ \bibinfo {pages} {43} (\bibinfo
  {year} {2012})},\ \Eprint {http://arxiv.org/abs/1106.1839} {arXiv:1106.1839
  [astro-ph.IM]} \BibitemShut {NoStop}%
\bibitem [{\citenamefont {Lutz}\ \emph {et~al.}(2016)\citenamefont {Lutz},
  \citenamefont {Porro}, \citenamefont {Aschauer}, \citenamefont {W{\"o}lfel},\
  and\ \citenamefont {Str{\"u}der}}]{Lutz:2016}%
  \BibitemOpen
  \bibfield  {author} {\bibinfo {author} {\bibfnamefont {G.}~\bibnamefont
  {Lutz}}, \bibinfo {author} {\bibfnamefont {M.}~\bibnamefont {Porro}},
  \bibinfo {author} {\bibfnamefont {S.}~\bibnamefont {Aschauer}}, \bibinfo
  {author} {\bibfnamefont {S.}~\bibnamefont {W{\"o}lfel}}, \ and\ \bibinfo
  {author} {\bibfnamefont {L.}~\bibnamefont {Str{\"u}der}},\ }\href {\doibase
  10.3390/s16050608} {\bibfield  {journal} {\bibinfo  {journal} {Sensors}\
  }\textbf {\bibinfo {volume} {16}},\ \bibinfo {pages} {608} (\bibinfo {year}
  {2016})}\BibitemShut {NoStop}%
\bibitem [{\citenamefont {Aguilar-Arevalo}\ \emph
  {et~al.}(2016{\natexlab{a}})\citenamefont {Aguilar-Arevalo} \emph
  {et~al.}}]{Aguilar-Arevalo:2016zop}%
  \BibitemOpen
  \bibfield  {author} {\bibinfo {author} {\bibfnamefont {A.}~\bibnamefont
  {Aguilar-Arevalo}} \emph {et~al.} (\bibinfo {collaboration} {DAMIC}),\
  }\href@noop {} {\bibfield  {journal} {\bibinfo  {journal} {Submitted to:
  Phys. Rev. Lett.}\ } (\bibinfo {year} {2016}{\natexlab{a}})},\ \Eprint
  {http://arxiv.org/abs/1611.03066} {arXiv:1611.03066 [astro-ph.CO]}
  \BibitemShut {NoStop}%
\bibitem [{\citenamefont {Holland}(1989)}]{Holland:1989}%
  \BibitemOpen
  \bibfield  {author} {\bibinfo {author} {\bibfnamefont {S.}~\bibnamefont
  {Holland}},\ }\href {\doibase http://dx.doi.org/10.1016/0168-9002(89)90741-9}
  {\bibfield  {journal} {\bibinfo  {journal} {Nuclear Instruments and Methods
  in Physics Research Section A: Accelerators, Spectrometers, Detectors and
  Associated Equipment}\ }\textbf {\bibinfo {volume} {275}},\ \bibinfo {pages}
  {537 } (\bibinfo {year} {1989})}\BibitemShut {NoStop}%
\bibitem [{\citenamefont {Holland}\ \emph {et~al.}(2003)\citenamefont
  {Holland}, \citenamefont {Groom}, \citenamefont {Palaio}, \citenamefont
  {Stover},\ and\ \citenamefont {Wei}}]{Holland:2003}%
  \BibitemOpen
  \bibfield  {author} {\bibinfo {author} {\bibfnamefont {S.~E.}\ \bibnamefont
  {Holland}}, \bibinfo {author} {\bibfnamefont {D.~E.}\ \bibnamefont {Groom}},
  \bibinfo {author} {\bibfnamefont {N.~P.}\ \bibnamefont {Palaio}}, \bibinfo
  {author} {\bibfnamefont {R.~J.}\ \bibnamefont {Stover}}, \ and\ \bibinfo
  {author} {\bibfnamefont {M.}~\bibnamefont {Wei}},\ }\href {\doibase
  10.1109/TED.2002.806476} {\bibfield  {journal} {\bibinfo  {journal} {IEEE
  Transactions on Electron Devices}\ }\textbf {\bibinfo {volume} {50}},\
  \bibinfo {pages} {225} (\bibinfo {year} {2003})}\BibitemShut {NoStop}%
\bibitem [{\citenamefont {Mclean}\ \emph {et~al.}(2012)\citenamefont {Mclean},
  \citenamefont {Flaugher}, \citenamefont {Abbott}, \citenamefont {Angstadt},
  \citenamefont {Annis} \emph {et~al.}}]{Mclean:2012pka}%
  \BibitemOpen
  \bibfield  {author} {\bibinfo {author} {\bibfnamefont {I.~S.}\ \bibnamefont
  {Mclean}}, \bibinfo {author} {\bibfnamefont {B.~L.}\ \bibnamefont
  {Flaugher}}, \bibinfo {author} {\bibfnamefont {T.~M.}\ \bibnamefont
  {Abbott}}, \bibinfo {author} {\bibfnamefont {R.}~\bibnamefont {Angstadt}},
  \bibinfo {author} {\bibfnamefont {J.}~\bibnamefont {Annis}},  \emph
  {et~al.},\ }\href {\doibase 10.1117/12.926216} {\bibfield  {journal}
  {\bibinfo  {journal} {Ground-based and Airborne Instrumentation for Astronomy
  IV}\ }\textbf {\bibinfo {volume} {8446}},\ \bibinfo {pages} {844611}
  (\bibinfo {year} {2012})}\BibitemShut {NoStop}%
\bibitem [{\citenamefont {Flaugher}\ \emph {et~al.}(2015)\citenamefont
  {Flaugher} \emph {et~al.}}]{Flaugher:2015}%
  \BibitemOpen
  \bibfield  {author} {\bibinfo {author} {\bibfnamefont {B.}~\bibnamefont
  {Flaugher}} \emph {et~al.} (\bibinfo {collaboration} {DES}),\ }\href
  {\doibase 10.1088/0004-6256/150/5/150} {\bibfield  {journal} {\bibinfo
  {journal} {Astron. J.}\ }\textbf {\bibinfo {volume} {150}},\ \bibinfo {pages}
  {150} (\bibinfo {year} {2015})},\ \Eprint {http://arxiv.org/abs/1504.02900}
  {arXiv:1504.02900 [astro-ph.IM]} \BibitemShut {NoStop}%
\bibitem [{\citenamefont {Haro}\ \emph {et~al.}(2016)\citenamefont {Haro},
  \citenamefont {Cancelo}, \citenamefont {Moroni}, \citenamefont {Bertou},
  \citenamefont {Tiffenberg}, \citenamefont {Paolini},\ and\ \citenamefont
  {Estrada}}]{Haro:2016}%
  \BibitemOpen
  \bibfield  {author} {\bibinfo {author} {\bibfnamefont {M.~S.}\ \bibnamefont
  {Haro}}, \bibinfo {author} {\bibfnamefont {G.}~\bibnamefont {Cancelo}},
  \bibinfo {author} {\bibfnamefont {G.~F.}\ \bibnamefont {Moroni}}, \bibinfo
  {author} {\bibfnamefont {X.}~\bibnamefont {Bertou}}, \bibinfo {author}
  {\bibfnamefont {J.}~\bibnamefont {Tiffenberg}}, \bibinfo {author}
  {\bibfnamefont {E.}~\bibnamefont {Paolini}}, \ and\ \bibinfo {author}
  {\bibfnamefont {J.}~\bibnamefont {Estrada}},\ }in\ \href {\doibase
  10.1109/CAMTA.2016.7574083} {\emph {\bibinfo {booktitle} {2016 Argentine
  Conference of Micro-Nanoelectronics, Technology and Applications (CAMTA)}}}\
  (\bibinfo {year} {2016})\ pp.\ \bibinfo {pages} {11--16}\BibitemShut
  {NoStop}%
\bibitem [{\citenamefont {O'Connor}(2015)}]{LSST:2015}%
  \BibitemOpen
  \bibfield  {author} {\bibinfo {author} {\bibfnamefont {P.}~\bibnamefont
  {O'Connor}},\ }\href {http://stacks.iop.org/1748-0221/10/i=05/a=C05010}
  {\bibfield  {journal} {\bibinfo  {journal} {Journal of Instrumentation}\
  }\textbf {\bibinfo {volume} {10}},\ \bibinfo {pages} {C05010} (\bibinfo
  {year} {2015})}\BibitemShut {NoStop}%
\bibitem [{\citenamefont {Doering}\ \emph {et~al.}(2011)\citenamefont
  {Doering}, \citenamefont {Andresen}, \citenamefont {Contarato}, \citenamefont
  {Denes}, \citenamefont {Joseph}, \citenamefont {McVittie}, \citenamefont
  {Walder}, \citenamefont {Weizeorick},\ and\ \citenamefont
  {Zheng}}]{Doering:2011}%
  \BibitemOpen
  \bibfield  {author} {\bibinfo {author} {\bibfnamefont {D.}~\bibnamefont
  {Doering}}, \bibinfo {author} {\bibfnamefont {N.}~\bibnamefont {Andresen}},
  \bibinfo {author} {\bibfnamefont {D.}~\bibnamefont {Contarato}}, \bibinfo
  {author} {\bibfnamefont {P.}~\bibnamefont {Denes}}, \bibinfo {author}
  {\bibfnamefont {J.~M.}\ \bibnamefont {Joseph}}, \bibinfo {author}
  {\bibfnamefont {P.}~\bibnamefont {McVittie}}, \bibinfo {author}
  {\bibfnamefont {J.~P.}\ \bibnamefont {Walder}}, \bibinfo {author}
  {\bibfnamefont {J.}~\bibnamefont {Weizeorick}}, \ and\ \bibinfo {author}
  {\bibfnamefont {B.}~\bibnamefont {Zheng}},\ }in\ \href {\doibase
  10.1109/NSSMIC.2011.6154370} {\emph {\bibinfo {booktitle} {2011 IEEE Nuclear
  Science Symposium Conference Record}}}\ (\bibinfo {year} {2011})\ pp.\
  \bibinfo {pages} {1840--1845}\BibitemShut {NoStop}%
\bibitem [{\citenamefont {Essig}\ \emph {et~al.}(2016)\citenamefont {Essig},
  \citenamefont {Fernandez-Serra}, \citenamefont {Mardon}, \citenamefont
  {Soto}, \citenamefont {Volansky},\ and\ \citenamefont {Yu}}]{Essig:2015cda}%
  \BibitemOpen
  \bibfield  {author} {\bibinfo {author} {\bibfnamefont {R.}~\bibnamefont
  {Essig}}, \bibinfo {author} {\bibfnamefont {M.}~\bibnamefont
  {Fernandez-Serra}}, \bibinfo {author} {\bibfnamefont {J.}~\bibnamefont
  {Mardon}}, \bibinfo {author} {\bibfnamefont {A.}~\bibnamefont {Soto}},
  \bibinfo {author} {\bibfnamefont {T.}~\bibnamefont {Volansky}}, \ and\
  \bibinfo {author} {\bibfnamefont {T.-T.}\ \bibnamefont {Yu}},\ }\href
  {\doibase 10.1007/JHEP05(2016)046} {\bibfield  {journal} {\bibinfo  {journal}
  {JHEP}\ }\textbf {\bibinfo {volume} {05}},\ \bibinfo {pages} {046} (\bibinfo
  {year} {2016})},\ \Eprint {http://arxiv.org/abs/1509.01598} {arXiv:1509.01598
  [hep-ph]} \BibitemShut {NoStop}%
\bibitem [{\citenamefont {Essig}\ \emph
  {et~al.}(2012{\natexlab{a}})\citenamefont {Essig}, \citenamefont {Mardon},\
  and\ \citenamefont {Volansky}}]{Essig:2011nj}%
  \BibitemOpen
  \bibfield  {author} {\bibinfo {author} {\bibfnamefont {R.}~\bibnamefont
  {Essig}}, \bibinfo {author} {\bibfnamefont {J.}~\bibnamefont {Mardon}}, \
  and\ \bibinfo {author} {\bibfnamefont {T.}~\bibnamefont {Volansky}},\ }\href
  {\doibase 10.1103/PhysRevD.85.076007} {\bibfield  {journal} {\bibinfo
  {journal} {Phys. Rev.}\ }\textbf {\bibinfo {volume} {D85}},\ \bibinfo {pages}
  {076007} (\bibinfo {year} {2012}{\natexlab{a}})},\ \Eprint
  {http://arxiv.org/abs/1108.5383} {arXiv:1108.5383 [hep-ph]} \BibitemShut
  {NoStop}%
\bibitem [{\citenamefont {Graham}\ \emph {et~al.}(2012)\citenamefont {Graham},
  \citenamefont {Kaplan}, \citenamefont {Rajendran},\ and\ \citenamefont
  {Walters}}]{Graham:2012su}%
  \BibitemOpen
  \bibfield  {author} {\bibinfo {author} {\bibfnamefont {P.~W.}\ \bibnamefont
  {Graham}}, \bibinfo {author} {\bibfnamefont {D.~E.}\ \bibnamefont {Kaplan}},
  \bibinfo {author} {\bibfnamefont {S.}~\bibnamefont {Rajendran}}, \ and\
  \bibinfo {author} {\bibfnamefont {M.~T.}\ \bibnamefont {Walters}},\ }\href
  {\doibase 10.1016/j.dark.2012.09.001} {\bibfield  {journal} {\bibinfo
  {journal} {Phys. Dark Univ.}\ }\textbf {\bibinfo {volume} {1}},\ \bibinfo
  {pages} {32} (\bibinfo {year} {2012})},\ \Eprint
  {http://arxiv.org/abs/1203.2531} {arXiv:1203.2531 [hep-ph]} \BibitemShut
  {NoStop}%
\bibitem [{\citenamefont {Lee}\ \emph {et~al.}(2015)\citenamefont {Lee},
  \citenamefont {Lisanti}, \citenamefont {Mishra-Sharma},\ and\ \citenamefont
  {Safdi}}]{Lee:2015qva}%
  \BibitemOpen
  \bibfield  {author} {\bibinfo {author} {\bibfnamefont {S.~K.}\ \bibnamefont
  {Lee}}, \bibinfo {author} {\bibfnamefont {M.}~\bibnamefont {Lisanti}},
  \bibinfo {author} {\bibfnamefont {S.}~\bibnamefont {Mishra-Sharma}}, \ and\
  \bibinfo {author} {\bibfnamefont {B.~R.}\ \bibnamefont {Safdi}},\ }\href
  {\doibase 10.1103/PhysRevD.92.083517} {\bibfield  {journal} {\bibinfo
  {journal} {Phys. Rev.}\ }\textbf {\bibinfo {volume} {D92}},\ \bibinfo {pages}
  {083517} (\bibinfo {year} {2015})},\ \Eprint
  {http://arxiv.org/abs/1508.07361} {arXiv:1508.07361 [hep-ph]} \BibitemShut
  {NoStop}%
\bibitem [{\citenamefont {An}\ \emph {et~al.}(2013)\citenamefont {An},
  \citenamefont {Pospelov},\ and\ \citenamefont {Pradler}}]{An:2013yua}%
  \BibitemOpen
  \bibfield  {author} {\bibinfo {author} {\bibfnamefont {H.}~\bibnamefont
  {An}}, \bibinfo {author} {\bibfnamefont {M.}~\bibnamefont {Pospelov}}, \ and\
  \bibinfo {author} {\bibfnamefont {J.}~\bibnamefont {Pradler}},\ }\href
  {\doibase 10.1103/PhysRevLett.111.041302} {\bibfield  {journal} {\bibinfo
  {journal} {Phys. Rev. Lett.}\ }\textbf {\bibinfo {volume} {111}},\ \bibinfo
  {pages} {041302} (\bibinfo {year} {2013})},\ \Eprint
  {http://arxiv.org/abs/1304.3461} {arXiv:1304.3461 [hep-ph]} \BibitemShut
  {NoStop}%
\bibitem [{\citenamefont {An}\ \emph {et~al.}(2015)\citenamefont {An},
  \citenamefont {Pospelov}, \citenamefont {Pradler},\ and\ \citenamefont
  {Ritz}}]{An:2014twa}%
  \BibitemOpen
  \bibfield  {author} {\bibinfo {author} {\bibfnamefont {H.}~\bibnamefont
  {An}}, \bibinfo {author} {\bibfnamefont {M.}~\bibnamefont {Pospelov}},
  \bibinfo {author} {\bibfnamefont {J.}~\bibnamefont {Pradler}}, \ and\
  \bibinfo {author} {\bibfnamefont {A.}~\bibnamefont {Ritz}},\ }\href {\doibase
  10.1016/j.physletb.2015.06.018} {\bibfield  {journal} {\bibinfo  {journal}
  {Phys. Lett.}\ }\textbf {\bibinfo {volume} {B747}},\ \bibinfo {pages} {331}
  (\bibinfo {year} {2015})},\ \Eprint {http://arxiv.org/abs/1412.8378}
  {arXiv:1412.8378 [hep-ph]} \BibitemShut {NoStop}%
\bibitem [{\citenamefont {Bloch}\ \emph {et~al.}(2016)\citenamefont {Bloch},
  \citenamefont {Essig}, \citenamefont {Tobioka}, \citenamefont {Volansky},\
  and\ \citenamefont {Yu}}]{Bloch:2016sjj}%
  \BibitemOpen
  \bibfield  {author} {\bibinfo {author} {\bibfnamefont {I.~M.}\ \bibnamefont
  {Bloch}}, \bibinfo {author} {\bibfnamefont {R.}~\bibnamefont {Essig}},
  \bibinfo {author} {\bibfnamefont {K.}~\bibnamefont {Tobioka}}, \bibinfo
  {author} {\bibfnamefont {T.}~\bibnamefont {Volansky}}, \ and\ \bibinfo
  {author} {\bibfnamefont {T.-T.}\ \bibnamefont {Yu}},\ }\href@noop {} {\
  (\bibinfo {year} {2016})},\ \Eprint {http://arxiv.org/abs/1608.02123}
  {arXiv:1608.02123 [hep-ph]} \BibitemShut {NoStop}%
\bibitem [{\citenamefont {Hochberg}\ \emph {et~al.}(2017)\citenamefont
  {Hochberg}, \citenamefont {Lin},\ and\ \citenamefont
  {Zurek}}]{Hochberg:2016sqx}%
  \BibitemOpen
  \bibfield  {author} {\bibinfo {author} {\bibfnamefont {Y.}~\bibnamefont
  {Hochberg}}, \bibinfo {author} {\bibfnamefont {T.}~\bibnamefont {Lin}}, \
  and\ \bibinfo {author} {\bibfnamefont {K.~M.}\ \bibnamefont {Zurek}},\ }\href
  {\doibase 10.1103/PhysRevD.95.023013} {\bibfield  {journal} {\bibinfo
  {journal} {Phys. Rev.}\ }\textbf {\bibinfo {volume} {D95}},\ \bibinfo {pages}
  {023013} (\bibinfo {year} {2017})},\ \Eprint
  {http://arxiv.org/abs/1608.01994} {arXiv:1608.01994 [hep-ph]} \BibitemShut
  {NoStop}%
\bibitem [{\citenamefont {Kouvaris}\ and\ \citenamefont
  {Pradler}(2017)}]{Kouvaris:2016afs}%
  \BibitemOpen
  \bibfield  {author} {\bibinfo {author} {\bibfnamefont {C.}~\bibnamefont
  {Kouvaris}}\ and\ \bibinfo {author} {\bibfnamefont {J.}~\bibnamefont
  {Pradler}},\ }\href {\doibase 10.1103/PhysRevLett.118.031803} {\bibfield
  {journal} {\bibinfo  {journal} {Phys. Rev. Lett.}\ }\textbf {\bibinfo
  {volume} {118}},\ \bibinfo {pages} {031803} (\bibinfo {year} {2017})},\
  \Eprint {http://arxiv.org/abs/1607.01789} {arXiv:1607.01789 [hep-ph]}
  \BibitemShut {NoStop}%
\bibitem [{\citenamefont {{Klein}}(1968)}]{Klein:1968}%
  \BibitemOpen
  \bibfield  {author} {\bibinfo {author} {\bibfnamefont {C.~A.}\ \bibnamefont
  {{Klein}}},\ }\href {\doibase 10.1063/1.1656484} {\bibfield  {journal}
  {\bibinfo  {journal} {Journal of Applied Physics}\ }\textbf {\bibinfo
  {volume} {39}},\ \bibinfo {pages} {2029} (\bibinfo {year}
  {1968})}\BibitemShut {NoStop}%
\bibitem [{\citenamefont {Streetman}(2005)}]{ExptGaps}%
  \BibitemOpen
  \bibfield  {author} {\bibinfo {author} {\bibfnamefont {B.~G. S.~B.}\
  \bibnamefont {Streetman}},\ }\href@noop {} {\bibfield  {journal} {\bibinfo
  {journal} {Prentice Hall}\ } (\bibinfo {year} {2005})}\BibitemShut {NoStop}%
\bibitem [{\citenamefont {Barreto}\ \emph {et~al.}(2012)\citenamefont {Barreto}
  \emph {et~al.}}]{Barreto:2011zu}%
  \BibitemOpen
  \bibfield  {author} {\bibinfo {author} {\bibfnamefont {J.}~\bibnamefont
  {Barreto}} \emph {et~al.} (\bibinfo {collaboration} {DAMIC Collaboration}),\
  }\href {\doibase 10.1016/j.physletb.2012.04.006} {\bibfield  {journal}
  {\bibinfo  {journal} {Phys.Lett.}\ }\textbf {\bibinfo {volume} {B711}},\
  \bibinfo {pages} {264} (\bibinfo {year} {2012})},\ \Eprint
  {http://arxiv.org/abs/1105.5191} {arXiv:1105.5191 [astro-ph.IM]} \BibitemShut
  {NoStop}%
\bibitem [{\citenamefont {Angle}\ \emph {et~al.}(2011)\citenamefont {Angle}
  \emph {et~al.}}]{Angle:2011th}%
  \BibitemOpen
  \bibfield  {author} {\bibinfo {author} {\bibfnamefont {J.}~\bibnamefont
  {Angle}} \emph {et~al.} (\bibinfo {collaboration} {XENON10}),\ }\href
  {\doibase 10.1103/PhysRevLett.110.249901, 10.1103/PhysRevLett.107.051301}
  {\bibfield  {journal} {\bibinfo  {journal} {Phys. Rev. Lett.}\ }\textbf
  {\bibinfo {volume} {107}},\ \bibinfo {pages} {051301} (\bibinfo {year}
  {2011})},\ \bibinfo {note} {[Erratum: Phys. Rev. Lett.110,249901(2013)]},\
  \Eprint {http://arxiv.org/abs/1104.3088} {arXiv:1104.3088 [astro-ph.CO]}
  \BibitemShut {NoStop}%
\bibitem [{\citenamefont {Essig}\ \emph
  {et~al.}(2012{\natexlab{b}})\citenamefont {Essig}, \citenamefont
  {Manalaysay}, \citenamefont {Mardon}, \citenamefont {Sorensen},\ and\
  \citenamefont {Volansky}}]{Essig:2012yx}%
  \BibitemOpen
  \bibfield  {author} {\bibinfo {author} {\bibfnamefont {R.}~\bibnamefont
  {Essig}}, \bibinfo {author} {\bibfnamefont {A.}~\bibnamefont {Manalaysay}},
  \bibinfo {author} {\bibfnamefont {J.}~\bibnamefont {Mardon}}, \bibinfo
  {author} {\bibfnamefont {P.}~\bibnamefont {Sorensen}}, \ and\ \bibinfo
  {author} {\bibfnamefont {T.}~\bibnamefont {Volansky}},\ }\href {\doibase
  10.1103/PhysRevLett.109.021301} {\bibfield  {journal} {\bibinfo  {journal}
  {Phys. Rev. Lett.}\ }\textbf {\bibinfo {volume} {109}},\ \bibinfo {pages}
  {021301} (\bibinfo {year} {2012}{\natexlab{b}})},\ \Eprint
  {http://arxiv.org/abs/1206.2644} {arXiv:1206.2644 [astro-ph.CO]} \BibitemShut
  {NoStop}%
\bibitem [{\citenamefont {Essig}\ \emph {et~al.}(2017)\citenamefont {Essig},
  \citenamefont {Volansky},\ and\ \citenamefont {Yu}}]{Essig:2017kqs}%
  \BibitemOpen
  \bibfield  {author} {\bibinfo {author} {\bibfnamefont {R.}~\bibnamefont
  {Essig}}, \bibinfo {author} {\bibfnamefont {T.}~\bibnamefont {Volansky}}, \
  and\ \bibinfo {author} {\bibfnamefont {T.-T.}\ \bibnamefont {Yu}},\
  }\href@noop {} {\  (\bibinfo {year} {2017})},\ \Eprint
  {http://arxiv.org/abs/1703.00910} {arXiv:1703.00910 [hep-ph]} \BibitemShut
  {NoStop}%
\bibitem [{\citenamefont {Sorensen}(2017)}]{Sorensen:2017ymt}%
  \BibitemOpen
  \bibfield  {author} {\bibinfo {author} {\bibfnamefont {P.}~\bibnamefont
  {Sorensen}},\ }\href@noop {} {\  (\bibinfo {year} {2017})},\ \Eprint
  {http://arxiv.org/abs/1702.04805} {arXiv:1702.04805 [physics.ins-det]}
  \BibitemShut {NoStop}%
\bibitem [{\citenamefont {Agnese}\ \emph {et~al.}(2016)\citenamefont {Agnese}
  \emph {et~al.}}]{Agnese:2015nto}%
  \BibitemOpen
  \bibfield  {author} {\bibinfo {author} {\bibfnamefont {R.}~\bibnamefont
  {Agnese}} \emph {et~al.} (\bibinfo {collaboration} {SuperCDMS}),\ }\href
  {\doibase 10.1103/PhysRevLett.116.071301} {\bibfield  {journal} {\bibinfo
  {journal} {Phys. Rev. Lett.}\ }\textbf {\bibinfo {volume} {116}},\ \bibinfo
  {pages} {071301} (\bibinfo {year} {2016})},\ \Eprint
  {http://arxiv.org/abs/1509.02448} {arXiv:1509.02448 [astro-ph.CO]}
  \BibitemShut {NoStop}%
\bibitem [{\citenamefont {Aalseth}\ \emph {et~al.}(2013)\citenamefont {Aalseth}
  \emph {et~al.}}]{Aalseth:2012if}%
  \BibitemOpen
  \bibfield  {author} {\bibinfo {author} {\bibfnamefont {C.~E.}\ \bibnamefont
  {Aalseth}} \emph {et~al.} (\bibinfo {collaboration} {CoGeNT}),\ }\href
  {\doibase 10.1103/PhysRevD.88.012002} {\bibfield  {journal} {\bibinfo
  {journal} {Phys. Rev.}\ }\textbf {\bibinfo {volume} {D88}},\ \bibinfo {pages}
  {012002} (\bibinfo {year} {2013})},\ \Eprint {http://arxiv.org/abs/1208.5737}
  {arXiv:1208.5737 [astro-ph.CO]} \BibitemShut {NoStop}%
\bibitem [{\citenamefont {Abgrall}\ \emph {et~al.}(2017)\citenamefont {Abgrall}
  \emph {et~al.}}]{Abgrall:2016tnn}%
  \BibitemOpen
  \bibfield  {author} {\bibinfo {author} {\bibfnamefont {N.}~\bibnamefont
  {Abgrall}} \emph {et~al.} (\bibinfo {collaboration} {Majorana}),\ }\href
  {\doibase 10.1103/PhysRevLett.118.161801} {\bibfield  {journal} {\bibinfo
  {journal} {Phys. Rev. Lett.}\ }\textbf {\bibinfo {volume} {118}},\ \bibinfo
  {pages} {161801} (\bibinfo {year} {2017})},\ \Eprint
  {http://arxiv.org/abs/1612.00886} {arXiv:1612.00886 [nucl-ex]} \BibitemShut
  {NoStop}%
\bibitem [{\citenamefont {Armengaud}\ \emph {et~al.}(2017)\citenamefont
  {Armengaud} \emph {et~al.}}]{Armengaud:2016aoz}%
  \BibitemOpen
  \bibfield  {author} {\bibinfo {author} {\bibfnamefont {E.}~\bibnamefont
  {Armengaud}} \emph {et~al.} (\bibinfo {collaboration} {EDELWEISS}),\ }\href
  {\doibase 10.1016/j.astropartphys.2017.03.006} {\bibfield  {journal}
  {\bibinfo  {journal} {Astropart. Phys.}\ }\textbf {\bibinfo {volume} {91}},\
  \bibinfo {pages} {51} (\bibinfo {year} {2017})},\ \Eprint
  {http://arxiv.org/abs/1607.04560} {arXiv:1607.04560 [astro-ph.CO]}
  \BibitemShut {NoStop}%
\bibitem [{\citenamefont {Fernandez~Moroni}\ \emph {et~al.}(2015)\citenamefont
  {Fernandez~Moroni}, \citenamefont {Estrada}, \citenamefont {Paolini},
  \citenamefont {Cancelo}, \citenamefont {Tiffenberg},\ and\ \citenamefont
  {Molina}}]{Moroni:2014wia}%
  \BibitemOpen
  \bibfield  {author} {\bibinfo {author} {\bibfnamefont {G.}~\bibnamefont
  {Fernandez~Moroni}}, \bibinfo {author} {\bibfnamefont {J.}~\bibnamefont
  {Estrada}}, \bibinfo {author} {\bibfnamefont {E.~E.}\ \bibnamefont
  {Paolini}}, \bibinfo {author} {\bibfnamefont {G.}~\bibnamefont {Cancelo}},
  \bibinfo {author} {\bibfnamefont {J.}~\bibnamefont {Tiffenberg}}, \ and\
  \bibinfo {author} {\bibfnamefont {J.}~\bibnamefont {Molina}},\ }\href
  {\doibase 10.1103/PhysRevD.91.072001} {\bibfield  {journal} {\bibinfo
  {journal} {Phys. Rev.}\ }\textbf {\bibinfo {volume} {D91}},\ \bibinfo {pages}
  {072001} (\bibinfo {year} {2015})},\ \Eprint {http://arxiv.org/abs/1405.5761}
  {arXiv:1405.5761 [physics.ins-det]} \BibitemShut {NoStop}%
\bibitem [{\citenamefont {{DAMIC Collaboration}}(2013)}]{DAMIC:2013}%
  \BibitemOpen
  \bibfield  {author} {\bibinfo {author} {\bibnamefont {{DAMIC
  Collaboration}}},\ }\href@noop {} {\bibfield  {journal} {\bibinfo  {journal}
  {TAUP}\ } (\bibinfo {year} {2013})}\BibitemShut {NoStop}%
\bibitem [{\citenamefont {Aguilar-Arevalo}\ \emph
  {et~al.}(2016{\natexlab{b}})\citenamefont {Aguilar-Arevalo} \emph
  {et~al.}}]{Aguilar-Arevalo:2016qen}%
  \BibitemOpen
  \bibfield  {author} {\bibinfo {author} {\bibfnamefont {A.}~\bibnamefont
  {Aguilar-Arevalo}} \emph {et~al.} (\bibinfo {collaboration} {CONNIE}),\
  }\href {\doibase 10.1088/1748-0221/11/07/P07024} {\bibfield  {journal}
  {\bibinfo  {journal} {JINST}\ }\textbf {\bibinfo {volume} {11}},\ \bibinfo
  {pages} {P07024} (\bibinfo {year} {2016}{\natexlab{b}})},\ \Eprint
  {http://arxiv.org/abs/1604.01343} {arXiv:1604.01343 [physics.ins-det]}
  \BibitemShut {NoStop}%
\bibitem [{\citenamefont {Harnik}\ \emph {et~al.}(2012)\citenamefont {Harnik},
  \citenamefont {Kopp},\ and\ \citenamefont {Machado}}]{Harnik:2012ni}%
  \BibitemOpen
  \bibfield  {author} {\bibinfo {author} {\bibfnamefont {R.}~\bibnamefont
  {Harnik}}, \bibinfo {author} {\bibfnamefont {J.}~\bibnamefont {Kopp}}, \ and\
  \bibinfo {author} {\bibfnamefont {P.~A.~N.}\ \bibnamefont {Machado}},\ }\href
  {\doibase 10.1088/1475-7516/2012/07/026} {\bibfield  {journal} {\bibinfo
  {journal} {JCAP}\ }\textbf {\bibinfo {volume} {1207}},\ \bibinfo {pages}
  {026} (\bibinfo {year} {2012})},\ \Eprint {http://arxiv.org/abs/1202.6073}
  {arXiv:1202.6073 [hep-ph]} \BibitemShut {NoStop}%
\bibitem [{\citenamefont {{National Research
  Council}}(2010)}]{DecadalReview2010}%
  \BibitemOpen
  \bibfield  {author} {\bibinfo {author} {\bibnamefont {{National Research
  Council}}},\ }\href {\doibase 10.17226/12951} {\emph {\bibinfo {title} {New
  Worlds, New Horizons in Astronomy and Astrophysics}}}\ (\bibinfo  {publisher}
  {The National Academies Press},\ \bibinfo {year} {2010})\BibitemShut
  {NoStop}%
\bibitem [{\citenamefont {Crill}\ and\ \citenamefont
  {Siegler}(2017)}]{ExoPlanet:2017}%
  \BibitemOpen
  \bibfield  {author} {\bibinfo {author} {\bibfnamefont {B.}~\bibnamefont
  {Crill}}\ and\ \bibinfo {author} {\bibfnamefont {N.}~\bibnamefont
  {Siegler}},\ }\href
  {https://exoplanets.nasa.gov/system/internal_resources/details/original/449_2017_ExEP_Technology_Plan_Appendix_Rev_B_Cleared.pdf}
  {\  (\bibinfo {year} {2017})}\BibitemShut {NoStop}%
\bibitem [{\citenamefont {{Shaklan}}\ \emph {et~al.}(2013)\citenamefont
  {{Shaklan}}, \citenamefont {{Levine}}, \citenamefont {{Foote}}, \citenamefont
  {{Rodgers}}, \citenamefont {{Underhill}}, \citenamefont {{Marchen}},\ and\
  \citenamefont {{Klein}}}]{Shaklan:2013}%
  \BibitemOpen
  \bibfield  {author} {\bibinfo {author} {\bibfnamefont {S.}~\bibnamefont
  {{Shaklan}}}, \bibinfo {author} {\bibfnamefont {M.}~\bibnamefont {{Levine}}},
  \bibinfo {author} {\bibfnamefont {M.}~\bibnamefont {{Foote}}}, \bibinfo
  {author} {\bibfnamefont {M.}~\bibnamefont {{Rodgers}}}, \bibinfo {author}
  {\bibfnamefont {M.}~\bibnamefont {{Underhill}}}, \bibinfo {author}
  {\bibfnamefont {L.}~\bibnamefont {{Marchen}}}, \ and\ \bibinfo {author}
  {\bibfnamefont {D.}~\bibnamefont {{Klein}}},\ }in\ \href {\doibase
  10.1117/12.2024560} {\emph {\bibinfo {booktitle} {Techniques and
  Instrumentation for Detection of Exoplanets VI}}},\ \bibinfo {series} {Proc.
  SPIE}, Vol.\ \bibinfo {volume} {8864}\ (\bibinfo {year} {2013})\ p.\ \bibinfo
  {pages} {886415}\BibitemShut {NoStop}%
\end{thebibliography}%

\end{document}